\documentclass{article}

\usepackage{PRIMEarxiv}

\usepackage[utf8]{inputenc} 
\usepackage[T1]{fontenc}    
\usepackage{hyperref}       
\usepackage{url}            
\usepackage{booktabs}       
\usepackage{amsfonts}       
\usepackage{nicefrac}       
\usepackage{microtype}      
\usepackage{lipsum}
\usepackage{fancyhdr}       
\usepackage{graphicx}       
\usepackage{tikz}
\usepackage{tikz-3dplot}
\usepackage{comment}
\usepackage{pgfplots}
\usepackage{textcomp}
\usepackage{mathtools}
\DeclareUnicodeCharacter{2212}{−}
\usepgfplotslibrary{groupplots,dateplot}
\usetikzlibrary{patterns,shapes.arrows}
\pgfplotsset{compat=newest}
\def\BibTeX{{\rm B\kern-.05em{\sc i\kern-.025em b}\kern-.08em
    T\kern-.1667em\lower.7ex\hbox{E}\kern-.125emX}}
\newcommand{\sn}[2]{\ensuremath{{#1}\times 10^{#2}}}
\graphicspath{{media/}}     

\title{On the Modelling and Performance Analysis of Lower Layer Mobility in 5G-Advanced 
}

\author{
  Alperen Gündogan, Akın Badalıoğlu, Panagiotis Spapis, Ahmad Awada \\
  \textit{Standardization and Research Lab, Nokia}\\
  Munich, Germany \\
   \texttt{\{alperen.gundogan, panagiotis.spapis, ahmad.awada\}@nokia-bell-labs.com} \\
   \texttt{akin.badalioglu.ext@nokia.com}
}

\begin{document}
\maketitle

\begin{abstract}
One of upcoming mobility enhancements in 5G-Advanced networks is to execute handover based on Layer 1 (L1) measurements using the so called lower layer mobility procedure. In this paper, we provide a system model for lower layer mobility procedure and we evaluate it against existing higher layer mobility procedures, such as baseline and conditional handover, using system level simulations. The benefits and drawbacks of lower layer mobility procedure are analyzed and compared against higher layer handover mechanisms using the relevant mobility key performance indicators. It has been shown that lower layer mobility procedure outperforms the existing handover mechanisms with respect to radio communication reliability at the expense of higher number of handovers and ping-pongs. To tackle these drawbacks, additional filtering for the L1 measurements used in handover decision is introduced to reduce the fluctuations caused by fast fading and measurement errors. Moreover, lower layer mobility procedure is enhanced with dynamic switching mechanism enabling the UE to change cells without being reconfigured by the network. The evaluations have shown that the introduction of such techniques is beneficial in reducing the number of ping-pongs and signaling overhead at the expense of an increase in the delay to react to rapid signal degradation and resource reservation overhead, respectively.
\end{abstract}


\section{Introduction}
The enhancements targeting reduction of the service interruption time during mobility are one of the main objectives in 5G. In particular, small interruption time while a UE is moving from one serving cell coverage area to another cell i.e., non-serving cell, is extremely crucial for ultra-reliable low latency communication (URLLC) applications, e.g. smart factory, healthcare industry with a strict reliability requirement of 99.999\% \cite{3gpp38913}.

The first release of 5G has defined the baseline handover mechanism which allows the network to control handover decisions based on the measurement report received from the UE. The measurement report includes cell quality measurements for serving and neighboring cells. Once a handover decision is a made by the serving cell, the network sends a Radio Resource Control (RRC) reconfiguration message, i.e., handover (HO) command, to the UE to initiate handover. Upon receiving the HO command, the UE detaches from the current serving cell and initiates random access to the target cell indicated by the network. The performance of the baseline handover mechanism highly depends on the timing of measurement report transmission and/or HO command reception. During the handover procedure, the radio link qualities between UE - serving cell and UE - target cell shall be good enough \cite{umurphd} to guarantee successful reception of the handover command from the serving cell and random access to the target cell.

The conditional handover (CHO) is introduced in \cite{R21706489} to increase the robustness of the baseline handover by configuring UEs with a condition to execute autonomously the handover. Once the condition is satisfied, the UE does not need to send measurement report nor wait to receive HO command from the serving cell, it can rather directly initiate the handover to the target cell. Although, CHO can provide enhanced mobility robustness compared to baseline handover \cite{9824571}, it still suffers from the high interruption time during the handover execution. 

Dual Active Protocol Stack (DAPS) handover is another solution that has been proposed in Rel.16 to reduce the handover interruption time down to 2ms \cite{3gpp38133}. This is achieved by using a make-before-break mechanism where the UE keeps the radio communication with the serving cell while performing the random access to the target cell. However, DAPS HO is not supported in higher frequencies e.g., intra-FR2 handover, which leaves a gap to improve the handover interruption time reduction for FR2. 
On the other hand, beam mobility was firstly discussed in Rel.17 in the scope of further enhancements on MIMO for NR work item where inter cell beam management (ICBM) was introduced. In the scope of this work item, the discussion on cell mobility using L1 measurements was triggered considering the advantages of ICBM but eventually postponed for future releases \cite{RP193133} \cite{RP212535}.

As a follow-up to Rel. 17 discussion, one of the objectives of Rel.18 mobility enhancements work item is to facilitate cell change based on L1 measurements where the handover is triggered by Medium Access Control (MAC) using lower layer signaling. This new solution for 5G-Advanced network will be supported for FR1 and FR2 and is named L1/L2 inter-cell mobility which is referred in this sequel as Lower Layer Mobility (LLM). LLM aims to reduce interruption during handover execution compared to baseline HO and CHO by restricting the reconfiguration during the handover execution and introducing new features such as RACH-less handover. The estimated interruption time can be as low as $1$ ms which is a tremendous decrease compared to baseline HO and CHO where the handover interruption is assumed to be around 80 ms \cite{3gppevm}. As such, LLM is a significant enhancement, especially for services requiring very high reliability such as URLLC.

The outline of the paper is as follows. In section \ref{section:Lower Layer Mobility}, LLM procedure is described along with the additional enhancements. The modelling of LLM procedure is presented in section \ref{section:Simulation Scenario and Model}. The simulation setup and scenario are discussed in section~\ref{sec:scenario}. Lastly, the mobility performance of LLM is analysed and compared with the baseline HO and CHO in section \ref{section:Performance Evaluation}.


\section{Lower Layer Mobility}
\label{section:Lower Layer Mobility}
The section describes the LLM procedure along with the enhancements.

\subsection{LLM Procedure}
According to the objectives of \cite{rp221799}, LLM is applicable for CU-DU split architecture, where CU manages PDCP, RRC layer and DU controls the PHY, MAC and RLC layer of the protocol stack. Fig. \ref{fig:llm_signal} illustrates a potential implementation of the LLM considering the three different phases; handover preparation, execution and completion \cite{nokiawp}. 

\subsubsection{Handover Preparation}
\begin{figure}[!t]
\centering
\resizebox{0.50\textwidth}{!}{\input{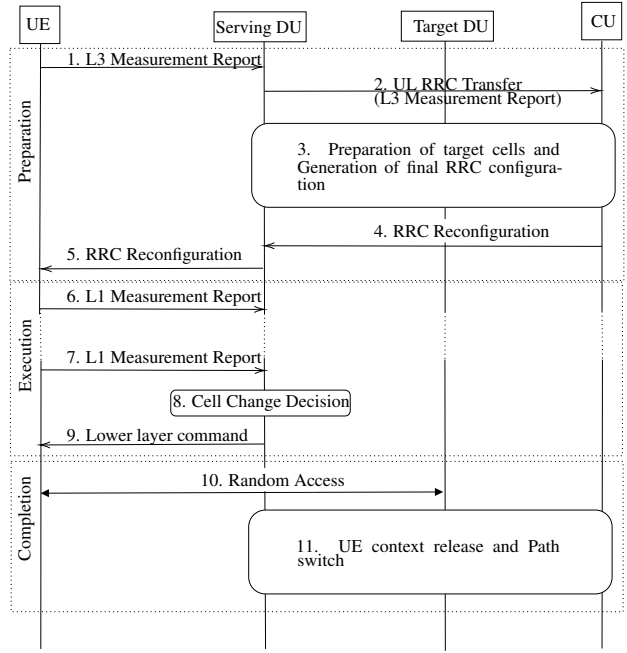}}%
\caption{High-level signalling diagram of lower layer mobility. \cite{nokiawp} }
\label{fig:llm_signal}
\end{figure}

The network provides the configurations of the potential target cells prior to the execution of LLM to enable pre-processing of configurations. The cell preparation for LLM is decided by CU based on the L3 measurement report that is received from the UE and which includes the cell quality measurements of the serving and neighboring cells. If the CU decides to initiate LLM, it requests from the relevant DUs to prepare configurations of the potential candidate cells for the UE. As shown in Fig.~\ref{fig:llm_signal}, the target cells may belong to the same DU as the serving cell (marked as Serving DU), or to another DU (marked as Target DU). The final configuration is sent to the UE in step 5 which contains the measurement configuration for reporting the L1 beam measurements of the prepared candidate cells and their configurations for handover.

\subsubsection{Handover Execution}
Upon receiving the configuration of the prepared target cells in step 5, UE starts to send L1 measurement report periodically to the network, including the L1 e.g., Reference Signal Received Power (RSRP) measurements of the serving and non-serving but prepared cell beams. The decision when to trigger the handover is left up to DU implementation. One straightforward approach is to trigger the handover when non-serving cell (NSC) beam signal strength becomes stronger than that of the serving cell (SC) by an offset, e.g. L1-RSRP of NSC $>$ L1-RSRP of SC + offset. Then the serving DU sends a lower layer command, e.g., MAC Control Element (CE) message, to trigger the cell change. 

\subsubsection{Handover Completion}
At step 10, UE detaches from the source cell and performs random access to the cell under the target DU to connect to the target cell. Upon successful RA procedure, UE context is released from the serving DU, and path switch is performed to the new target DU.

\subsection{LLM Enhancements}
To reduce the handover interruption time in LLM, RACH-less handover is one enhancement that is expected to be specified for LLM procedure. Herein, the random access to the target cell, shown in step 10 of Fig.~\ref{fig:llm_signal}, is skipped and the same timing advance of the source cell is re-used for the target cell. 

Another enhancement is the dynamic switching (DS) mechanism among the prepared candidate cells which is captured as one objective in 3GPP Rel. 18 Mobility Enhancements work item \cite{rp221799}. The aim of dynamic switching is to reduce the signaling overhead and to increase mobility robustness by allowing the UE to keep the prepared cell configurations after each cell change. Thus, the network does not need to reconfigure the target cells which were prepared before the handover. 


\section{Modelling of Lower Layer Mobility} 
\label{section:Simulation Scenario and Model}
This section explains the modelling for LLM procedure. 

\subsection{Cell Preparation, Release and Replacement}
In LLM mobility, the users are configured to perform RSRP measurements \( P_{c,b}^{\text{RSRP}} (t)\)  of synchronization signal block (SSB) for each beam \(b \in B\) of cell \(c \in C\)  at time \(t\) with periodicity of \( w \in  \mathbb{N} \). The raw measurements are filtered at the UE with a finite impulse response (FIR) filter in order to reduce the fluctuations of the signal caused by fast fading and measurement errors, i.e.,
\begin{equation}
\label{eq:l1filter}
		P_{c,b}^{\text{L1}} (t) =  \frac{1}{N_{L1}} \sum_{n=0}^{N_{L1}-1} P_{c,b}^{\text{RSRP}} (t -nw).
\end{equation}
The window size  \( {N_{L1}} \) used in Eq.\ref{eq:l1filter} is left for UE implementation according to the measurement accuracy requirements of user device \cite{3gpp38133}.

To derive L3 measurements for triggering the cell preparation in LLM, beam consolidation is performed first where the strongest L1 beam measurement in a cell, denoted by $P_{c}^{\text{L1}} (t) =\max_{b}  P_{c,b}^{\text{L1}} (t)$, is picked up at each time instant $t$. Then, L3 filtering with an infinite impulse response (IIR) filter is applied to smoothen the residual fluctuations in $P_{c}^{\text{L1}} (t)$ and increase its measurement accuracy, i.e.,
\begin{equation}
		P_{c}^{\text{L3}} (t) = \alpha P_{c}^{\text{L1}} (t) + (1-\alpha) P_{c}^{\text{L3}} (t-w),
\end{equation}where \( \alpha = (\frac{1}{2})^{k_{\text{L3}}/4}\) is the forgetting factor which controls the impact of the earlier measurements \(  P_{c}^{\text{L3}} (t-w)\) on the final value \(  P_{c}^{\text{L3}} (t)\) at time \( t \) and \( k_{\text{L3}} \) is the filter coefficient \cite{3gpp38331}. 

The UE sends the measurement report at step 1 of Fig. \ref{fig:llm_signal} if LLM preparation condition based on the L3 filtered RSRP values is satisfied. The condition for cell preparation is similar to the A3 event where a target cell  $c'$ is prepared if its signal strength is not weaker than the serving cell signal strength $c$ by more than an offset $o_{c,c'}^{\text{prep}}$ during the monitoring window \( T_{\text{prep}}\). The cell preparation entering condition is defined as
\begin{equation}
	 o_{c,c'}^{\text{prep}} > P_{c}^{\text{L3}} (t)  - P_{c'}^{\text{L3}} (t)  \ \text{for}  \  t _{\text{prep}}-T_{\text{prep}} < t <  t _{\text{prep}},
\end{equation}
where \( o_{c,c'}^{\text{prep}} \) is the cell preparation offset and \(  P_{c'}^{L3} (t) \) is the measurement of the potential target cell $c'$. The UE sends measurement reports for cell preparation if this condition holds at time \(  t = t_{prep}\).

If the cell preparation leaving condition is satisfied, the UE sends again a measurement report to the serving cell to inform that the prepared cell \( c'\) is no longer relevant for LLM, such that the network can release the configuration of target cell $c'$ and its resources prepared for the UE. The leaving condition is defined as
\begin{equation} \label{eqn:prepleave}
	 o_{c,c'}^{\text{prep}} <  P_{c}^{\text{L3}} (t)  - P_{c'}^{\text{L3}} (t)  \ \text{for}  \  t _{\text{prep}}-T_{\text{prep}} < t <  t _{\text{prep}}.
\end{equation}

 The maximum number of cells prepared by the network is limited to \( L \in  \mathbb{N} \) in order to constrain the resource reservation in the network. If the list of prepared cells is full and another potential cell \( c''\) satisfies the cell preparation entering condition, the UE compares its \(  P_{c''}^{\text{L3}} (t) \) with that of the weakest cell among the prepared ones. If \(  P_{c''}^{\text{L3}} (t)  > \min_{c  \in (c_1, c_2,..., c_L)}   P_{c}^{\text{L3}} (t)  \)  is satisfied, the UE replaces the weakest cell in the prepared cell list with the new cell \( c''\).  
\subsection{Handover Execution}
Once the preparation phase is completed, the UE starts to report the L1-RSRP values, i.e. \( P_{c,b}^{\text{L1}} (t) \) of the beams that belong to the prepared cells. The strongest $K$ L1 RSRP measurements, \( 	P_{c,b}^{\text{L1}} (t) \) of both serving and prepared cells are included in the L1 measurement report. Using the received measurements, the network decides to execute the handover if \( P_{c',b}^{\text{L1}}\) of target cell $c'$ is stronger than \( P_{c,b}^{\text{L1}} \) of serving cell $c$ by an offset $o_{c,c'}^{\text{exec}}$.

\section{Simulation Scenario and Parameters}
\label{sec:scenario}

In this section, the simulation scenario and parameters that are used for performance evaluation are described.
\begin{figure}[h]
\centering
\includegraphics[width=8cm]{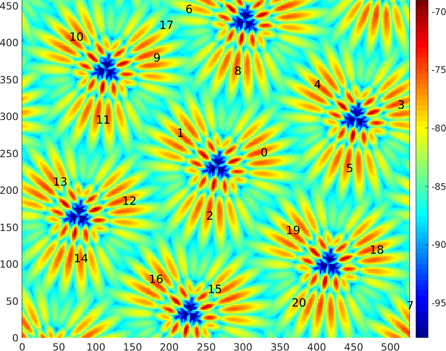}
\caption{Coverage map of urban macro scenario in FR2.}
\label{fig:coveragemap}
\end{figure}

\begin{table}[htbp]
\caption{Simulation Setup.}
\begin{center}
\begin{tabular}{|c|c|}
\hline
\textbf{Parameters} & \textbf{Values} \\
\hline
Environment & 3GPP compliant 7 base stations, \\ & 3 sector urban macro  scenario \\ & with 200 meters inter-site distance  \\ & (3D UMa channel model) \cite{3gpp38901} \\
\hline
Antenna Settings & 2x8x16 antenna panel with  14-beams  \\
\hline
Grid of Beams  & Azimuth angle = $ \{ 36.6^\circ, 50^\circ, 63.3^\circ, 76.6^\circ, 90^\circ, 103.3^\circ $ , \\ & $ 116.6^\circ,  130^\circ, 143.3^\circ, 42^\circ, 60^\circ, 90^\circ, 120^\circ, 138^\circ \} $ \\ & Elevation angle = \{ $-13^\circ, -10^\circ, -10^\circ, -10^\circ, -11^\circ, -10^\circ $, \\ & $ -10^\circ, -10^\circ, -13^\circ, -30^\circ, -33^\circ, -36^\circ, -33^\circ, -30^\circ$\}\\
\hline
PHY Settings & 28 GHz carrier frequency – one frequency layer \\ & only,  20.16 MHz bandwidth, 120 kHz \\ & subcarrier spacing, TDD (DSUDDDDDDD) \\
\hline
MAC Settings & Proportional fair in time and frequency \\ & scheduling  with TTI size of one subframe \\
\hline
Configurations & Simulation duration = 10 s, number of UEs = 420, \\ & speed of UEs = 60 km/h \\
\hline
Other & SSB transmission periodicity = 20 ms,
\\ & number of drops = 10, UL is disabled,
\\ & FR2 with analog beam-forming \\
\hline
\end{tabular}
\label{tab:simulation_setup}
\end{center}
\end{table}

A proprietary system-level simulator (SLS) is used for performance analyses of the different mobility procedures including baseline HO, CHO and LLM. The 3GPP complaint three sector urban macro scenario for a network in FR2 is adopted for performance evaluation. The network consists of 7 base stations (BSs), each serving three sectorized cells. Each cell is configured with 14 beams and transmits with a power $44~\text{dBm}$ in different azimuth angles. The system is kept in maximum load by using full-buffer traffic generator. The total 420 users are created randomly in the simulation and move along a direction with a speed of $60~\text{km/h}$. Further details of the simulation setup are shown in Table \ref{tab:simulation_setup}
including simulation environment, antenna settings, physical layer settings and MAC layer settings.
Fig. \ref{fig:coveragemap} shows the coverage map of the cell without slow and fast fading. For proper interference calculation at the outer cells, wrap around method has been used by wrapping the signals and mobility from others edges.

\section{Performance Evaluation}
\label{section:Performance Evaluation}
In this section, LLM and its corresponding enhancements are evaluated and compared against BHO and CHO. 

\subsection{Configuration of Mobility Procedures}
Since in CHO the RRC configurations of the target cells are also provided in advance, in order to have a fair comparison, two variants are evaluated: 1) The CHO execution condition is evaluated using L3 filtered cell quality measurements $P_{c}^{\text{L3}} (t)$ and TTT is set to $160~\text{ms}$ and 2) The CHO execution condition relies on L1 filtered cell RSRP measurements $	P_{c}^{\text{L1}} (t)$ and TTT is disabled by setting $0$ (denoted as CHO-L1). LLM is evaluated for three different cases: In case 1, denoted by LLM, the handover execution relies on L1 filtered beam RSRP measurement $	P_{c,b}^{\text{L1}} (t)$. In the second case, denoted by LLM-F, additional L2 filtering is applied to L1 filtered RSRP measurements. The L2 filter is an IIR filter which is similar to the L3 filter presented in Section \ref{section:Simulation Scenario and Model} and is applied at the serving cell to the reported L1-RSRP beam measurements. At last,  the impact of dynamic switching for LLM is investigated in case 3 which is denoted by LLM-F-DS. Herein, L2 filtering is enabled as in LLM-F but additionally the UE keeps the configurations of the prepared cells after the handover as long as the cell preparation leave event in Eq.~\ref{eqn:prepleave} or cell replacement event is not satisfied. The handover preparation and execution offsets are set in all the procedures to $3~\text{dB}$. 

The key parameters that distinguish the characteristics of mentioned mobility procedures can be found in Table \ref{tab:mob_procedure_params}. The handover interruption time is set to $80~\text{ms}$ and $1 ~\text{ms}$ for RRC mobility procedures, i.e. BHO/CHO  and LLM procedures, respectively, following 3GPP assumptions \cite{3gppevm}. The simulations are run in total for 10 drops with a random seed that affects mobility and radio propagation behaviour in order to obtain sufficient statistics in dynamic simulation environment.


\begin{table}[htbp]
\caption{Parameters of Mobility Procedures.}
\begin{center}
\begin{tabular}{|c|c|c|c|c|c|}
\hline
\textbf{Parameters}&\multicolumn{5}{|c|}{\textbf{Mobility Procedures}} \\
\cline{2-6} 
& \textbf{BHO} & \textbf{CHO}& \textbf{CHO-L1}& \textbf{LLM} & \textbf{LLM-F}, \textbf{LLM-F-DS}  \\
\hline
Handover  & 40 & 40 & 40 & 40 & 40 \\ Preparation & & & & & \\ Delay (ms) & & & & &\\
\hline
Handover & 80 & 80 & 80 & 1 & 1 \\ Interruption & & & & & \\ Time (ms) & & & & &\\
\hline
L2 Filter & 0 & 0 & 0 & 0 & 3 \\ Coefficient & & & & & \\
\hline
Handover Time & 160 & 160 & 0 & 0 & 0 \\ to Trigger (ms) & & & & & \\
\hline
Preparation & n/a & 3 & 3 & 3 & 3 \\ Offset (dB) & & & & & \\
\hline
Execution & 3 & 3 & 3 & 3 & 3 \\ Offset (dB) & & & & & \\
\hline
Max Number of & n/a & 4 & 4 & 4 & 4 \\ Prepared Cells & & & & & \\
\hline
\end{tabular}
\label{tab:mob_procedure_params}
\end{center}
\end{table}

\subsection{Mobility Performance Indicators}
In order to evaluate the mobility performance of LLM, BHO, and CHO, a set of mobility relevant key performance indicators (KPIs) are used which capture bad radio link quality, unnecessary handovers, and excessive resource reservation: 

1) Bad radio link quality is captured by \textbf{radio link problems (RLPs)} and \textbf{handover failures (HOFs)}. The former may occur upon receiving N310 consecutive out-of-sync indications from lower layers which indicates that signal quality between serving cell and the UE gets worse. The latter is declared if the random access to the target cell is not successful. This can indicate that the target cell signal at the time of HO execution is not stable enough for UE to connect to the target cell. 

2) Unnecessary handovers can be captured by measuring the \textbf{ping-pong (PP)} events, which are the cases that the handover to the target cell is successful but the UE performs another handover to the previous serving cell within $1000$ ms. Ping-pongs cause interruption during the handover and introduce signaling overhead and outage to the system. 

3) As outage can be caused either by failure or successful handover, we use \textbf{reliability}  in $\%$ to understand how reliable the overall communication is between the UE and the network. This statistic is based on the total service time (which is equal to the simulation time) and total time of outage considering all the UEs. The ratio of total time of outage to total service time provides the percentage of outage from which the reliability percentage can be derived by taking, i.e.,  100\%-outage [\%]. 

4) In addition, the number of \textbf{cell preparation} events for LLM procedures is analyzed in order to evaluate the signaling overhead on the radio interface between UE and the serving cell, and network signalling overhead. 

5) During the handover preparation, the target cell reserves resources for the UE including random access preambles, radio resources, buffers and UE identifier, etc. 
To capture this aspect, we define the metric of network \textbf{resource reservation} percentage for each UE in the network. This metric is calculated based on the time duration between a cell preparation and release for a UE:
Resource reservation periods of all cells are summed up for all the UEs and this sum is normalized by number of drops, number of UEs and simulation time to achieve the resource reservation percentage of the whole network for one UE.

\subsection{Performance Results}

\begin{figure*}[h]
\centering
    \resizebox{0.4\textwidth}{!}{
\begin{tikzpicture}
 \tikzstyle{every node}=[font=\small]
\definecolor{darkgray176}{RGB}{176,176,176}
\definecolor{green}{RGB}{0,128,0}
\definecolor{lightgray204}{RGB}{204,204,204}
\definecolor{magenta}{RGB}{255,0,255}
\definecolor{yellow}{RGB}{255,255,0}

\begin{axis}[
legend style={fill opacity=0.8, draw opacity=1, text opacity=1, draw=lightgray204},
tick align=outside,
tick pos=left,
x grid style={darkgray176},
xmajorgrids,
xmin=-0.69, xmax=16.29,
xtick={0.6, 3.4, 6.2, 9.0, 11.8, 15.0},
xticklabels={BHO,CHO,CHO-L1,LLM,LLM-F,LLM-F-DS},
xtick style={color=black},
y grid style={darkgray176},
ylabel={RLP/UE/min},
ymajorgrids,
ymin=0, ymax=0.733333333333333,
ytick style={color=black}
]
\draw[draw=none,fill=blue] (axis cs:-0.4,0) rectangle (axis cs:1.6,0.698412698412698);
\draw[draw=none,fill=yellow] (axis cs:2.4,0) rectangle (axis cs:4.4,0.538095238095238);
\draw[draw=none,fill=magenta] (axis cs:5.2,0) rectangle (axis cs:7.2,0.266666666666667);
\draw[draw=none,fill=red] (axis cs:8.0,0) rectangle (axis cs:10.0,0.2);
\draw[draw=none,fill=green] (axis cs:10.8,0) rectangle (axis cs:12.8,0.257142);
\draw[draw=none,fill=black] (axis cs:14.0,0) rectangle (axis cs:16.0, 0.19365);
\end{axis}

\end{tikzpicture}}\hfil
    \resizebox{0.4\textwidth}{!}{
\begin{tikzpicture}
 \tikzstyle{every node}=[font=\small]
\definecolor{darkgray176}{RGB}{176,176,176}
\definecolor{green}{RGB}{0,128,0}
\definecolor{lightgray204}{RGB}{204,204,204}
\definecolor{magenta}{RGB}{255,0,255}
\definecolor{yellow}{RGB}{255,255,0}

\begin{axis}[
legend style={fill opacity=0.8, draw opacity=1, text opacity=1, draw=lightgray204},
tick align=outside,
tick pos=left,
x grid style={darkgray176},
xmajorgrids,
xmin=-0.69, xmax=16.29,
xtick={0.6, 3.4, 6.2, 9.0, 11.8, 15.0},
xticklabels={BHO,CHO,CHO-L1,LLM,LLM-F,LLM-F-DS},
xtick style={color=black},
y grid style={darkgray176},
ylabel={HOF/UE/min},
ymajorgrids,
ymin=0, ymax=0.09999,
ytick style={color=black}
]
\draw[draw=none,fill=blue] (axis cs:-0.4,0) rectangle (axis cs:1.6,0.0714);
\draw[draw=none,fill=yellow] (axis cs:2.4,0) rectangle (axis cs:4.4,0.0428);
\draw[draw=none,fill=magenta] (axis cs:5.2,0) rectangle (axis cs:7.2,0.073);
\draw[draw=none,fill=red] (axis cs:8.0,0) rectangle (axis cs:10.0,0.0809);
\draw[draw=none,fill=green] (axis cs:10.8,0) rectangle (axis cs:12.8,0.0714);
\draw[draw=none,fill=black] (axis cs:14.0,0) rectangle (axis cs:16.0, 0.0984);
\end{axis}

\end{tikzpicture}}\par\medskip
    \resizebox{0.4\textwidth}{!}{
\begin{tikzpicture}
 \tikzstyle{every node}=[font=\small]
\definecolor{darkgray176}{RGB}{176,176,176}
\definecolor{green}{RGB}{0,128,0}
\definecolor{lightgray204}{RGB}{204,204,204}
\definecolor{magenta}{RGB}{255,0,255}
\definecolor{yellow}{RGB}{255,255,0}

\begin{axis}[
legend style={fill opacity=0.8, draw opacity=1, text opacity=1, draw=lightgray204},
tick align=outside,
tick pos=left,
x grid style={darkgray176},
xmajorgrids,
xmin=-0.69, xmax=16.29,
xtick={0.6, 3.4, 6.2, 9.0, 11.8, 15.0},
xticklabels={BHO,CHO,CHO-L1,LLM,LLM-F,LLM-F-DS},
xtick style={color=black},
y grid style={darkgray176},
ylabel={PP/UE/min},
ymajorgrids,
ymin=0, ymax=14.733333333333333,
ytick style={color=black}
]
\draw[draw=none,fill=blue] (axis cs:-0.4,0) rectangle (axis cs:1.6,3.90);
\draw[draw=none,fill=yellow] (axis cs:2.4,0) rectangle (axis cs:4.4,2.44);
\draw[draw=none,fill=magenta] (axis cs:5.2,0) rectangle (axis cs:7.2, 9.98);
\draw[draw=none,fill=red] (axis cs:8.0,0) rectangle (axis cs:10.0, 14.29);
\draw[draw=none,fill=green] (axis cs:10.8,0) rectangle (axis cs:12.8, 10.26);
\draw[draw=none,fill=black] (axis cs:14.0,0) rectangle (axis cs:16.0, 9.36);
\end{axis}

\end{tikzpicture}}\hfil
   \resizebox{0.4\textwidth}{!}{
\begin{tikzpicture}
 \tikzstyle{every node}=[font=\small]
\definecolor{darkgray176}{RGB}{176,176,176}
\definecolor{green}{RGB}{0,128,0}
\definecolor{lightgray204}{RGB}{204,204,204}
\definecolor{magenta}{RGB}{255,0,255}
\definecolor{yellow}{RGB}{255,255,0}

\begin{axis}[
legend style={fill opacity=0.8, draw opacity=1, text opacity=1, draw=lightgray204},
tick align=outside,
tick pos=left,
x grid style={darkgray176},
xmajorgrids,
xmin=-0.69, xmax=16.29,
xtick={0.6, 3.4, 6.2, 9.0, 11.8, 15.0},
xticklabels={BHO,CHO,CHO-L1,LLM,LLM-F,LLM-F-DS},
xtick style={color=black},
y grid style={darkgray176},
ylabel={Reliability [\%]},
ymajorgrids,
ymin=90, ymax=100,
ytick style={color=black}
]
\draw[draw=none,fill=blue] (axis cs:-0.4,0) rectangle (axis cs:1.6,95.58);
\draw[draw=none,fill=yellow] (axis cs:2.4,0) rectangle (axis cs:4.4,96.11);
\draw[draw=none,fill=magenta] (axis cs:5.2,0) rectangle (axis cs:7.2,94.64);
\draw[draw=none,fill=red] (axis cs:8.0,0) rectangle (axis cs:10.0,99.34);
\draw[draw=none,fill=green] (axis cs:10.8,0) rectangle (axis cs:12.8,99.25);
\draw[draw=none,fill=black] (axis cs:14.0,0) rectangle (axis cs:16.0, 99.21);
\end{axis}

\end{tikzpicture}}
\caption{The performance of different mobility procedures; radio link problems (top left), handover failures (top right), ping-pongs (bottom left), reliability (bottom-right)}
\label{fig:kpi_comparison}
\end{figure*}
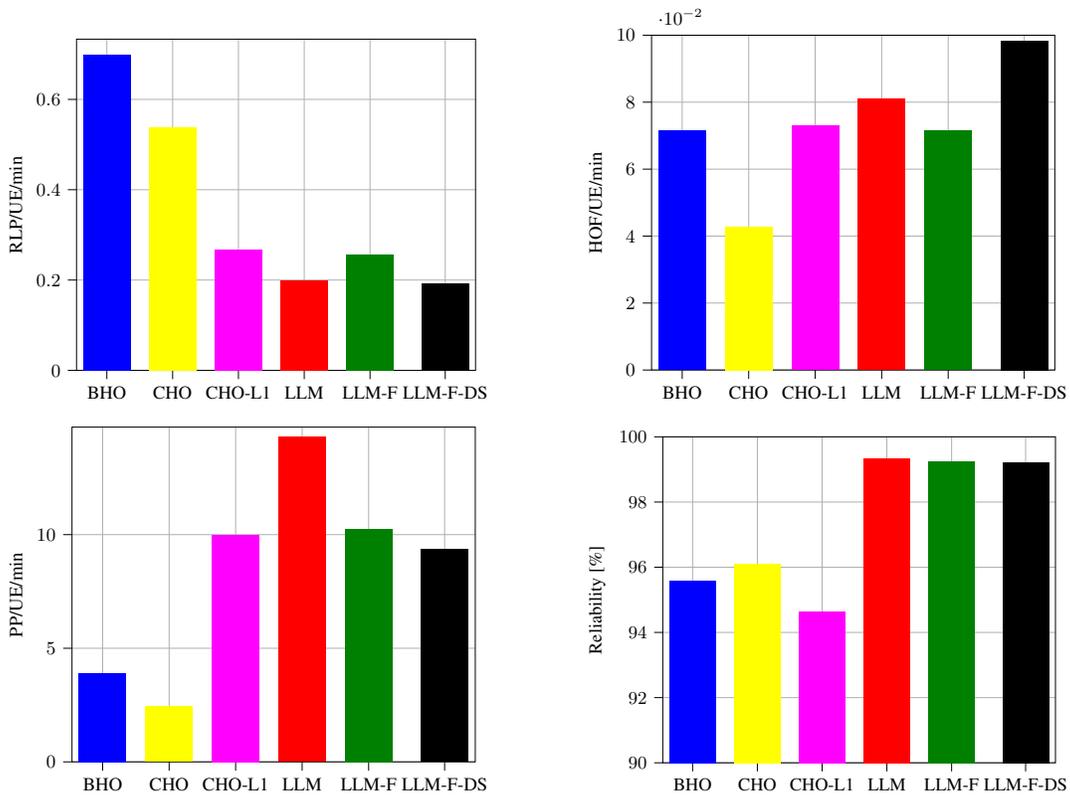

The numbers of RLP, HOF, PP are normalized per UE per minute and they are shown together with reliability metric in Fig. \ref{fig:kpi_comparison} as ``RLP/UE/min" (top left), ``HOF/UE/min"(top right), ``PP/UE/min" (bottom left), and reliability (bottom right), respectively. 

According to the Fig. \ref{fig:kpi_comparison}, LLM reduces the number of RLPs by around $71\%$ and $62\%$ against BHO and CHO, respectively. However, the number of HOFs for LLM is higher by around $13\%$ against BHO and $89\%$ against CHO. The high number of HOFs for LLM is caused by many PPs which indicates that handover execution in LLM may be triggered when the target cell is not strong enough for the UE. It should be noted though that the PPs in LLM are not costly in terms of handover interruption as it is $1/80$ of that in higher layer mobility procedures. Therefore, LLM outperforms BHO and CHO in terms of reliability by reaching around $99.3\%$ of reliability. 


Compared to CHO, the number of RLPs can be reduced by around $50\%$ in CHO-L1 as UE can avoid the L3 filtering delay and adapt faster to the channel degradations in serving cell. However, this increases substantially the number of PPs and therefore, the overall reliability of CHO-L1 is smaller than that of CHO as each PP costs at least $80~\text{ms}$ of interruption time.

In LLM, the number of PPs is decreased by $28\%$ when  additional IIR filtering is applied to the beam measurements as shown by LLM-F. However, filtering adds additional delay for handover execution which leads to an increase of $22\%$ in RLP compared to LLM. The reliability of LLM-F is only slightly reduced compared to LLM and is negligible. Thus, the signalling overhead in network and radio interface can be reduced by applying L2 filtering without degrading reliability.

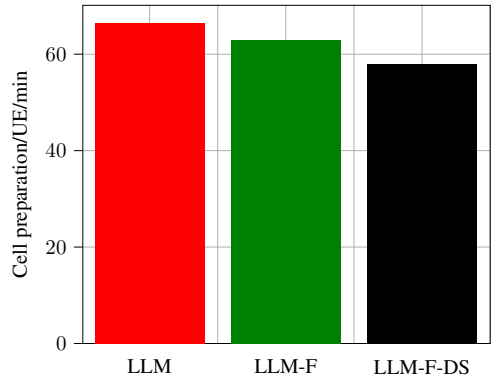
\begin{figure}[!t]
\centering
\resizebox{0.4\textwidth}{!}{

\begin{tikzpicture}

\definecolor{darkgray176}{RGB}{176,176,176}
\definecolor{green}{RGB}{0,128,0}

\begin{axis}[
tick align=outside,
tick pos=left,
x grid style={darkgray176},
xmajorgrids,
xmin=-0.49, xmax=2.5,
xtick style={draw=none},
xticklabels= {,LLM,,LLM-F,, LLM-F-DS}, 
y grid style={darkgray176},
ylabel={Cell preparation/UE/min},
ymajorgrids,
ymin=0, ymax=70.1333333333333,
ytick style={color=black}
]
\draw[draw=none,fill=red] (axis cs:-0.4,0) rectangle (axis cs:0.4,66.4269);
\draw[draw=none,fill=green] (axis cs:0.6,0) rectangle (axis cs:1.4,62.8476);
\draw[draw=none,fill=black] (axis cs:1.6,0) rectangle (axis cs:2.4,58.0222);
\end{axis}

\end{tikzpicture}}%
\caption{Number of cell preparation/UE/min for the three LLM cases.}
\label{fig:sentcellprep}
\end{figure}

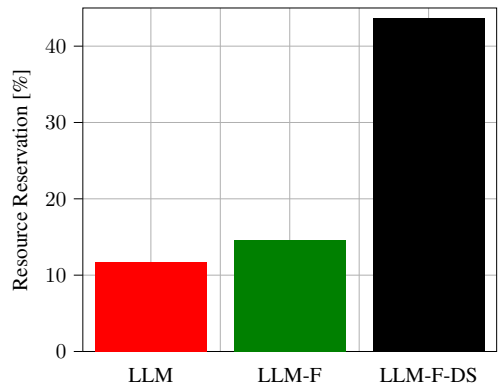
\begin{figure}[!t]
\centering
\resizebox{0.4\textwidth}{!}{
\begin{tikzpicture}

\definecolor{darkgray176}{RGB}{176,176,176}
\definecolor{green}{RGB}{0,128,0}

\begin{axis}[
tick align=outside,
tick pos=left,
x grid style={darkgray176},
xmajorgrids,
xmin=-0.49, xmax=2.49,
xtick style={draw=none},
xticklabels={,LLM,,LLM-F,,LLM-F-DS},
y grid style={darkgray176},
ylabel={Resource Reservation [\%]},
ymajorgrids,
ymin=0, ymax=45,
ytick style={color=black}
]
\draw[draw=none,fill=red] (axis cs:-0.4,0) rectangle (axis cs:0.4,11.76);
\draw[draw=none,fill=green] (axis cs:0.6,0) rectangle (axis cs:1.4,14.55);
\draw[draw=none,fill=black] (axis cs:1.6,0) rectangle (axis cs:2.4,43.64);
\end{axis}

\end{tikzpicture}}%
\caption{Resource reservation percentage for a UE in the network for the three LLM cases.}
\label{fig:resourcereservation_ds_llm}
\end{figure}

\begin{figure}[!t]
\centering
\resizebox{0.4\textwidth}{!}{\input{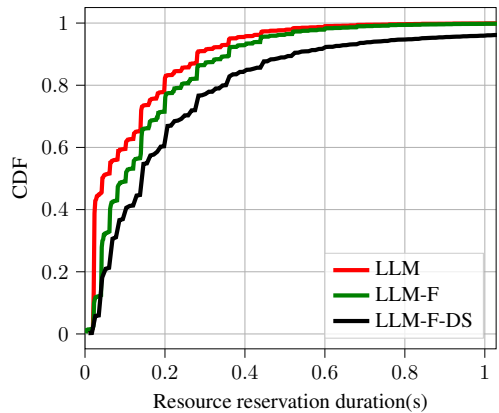}}%
\caption{CDF of resource reservation duration in a prepared cell.}
\label{fig:resourcereservation_ds_cdf}
\end{figure}

Fig.~\ref{fig:sentcellprep} shows the number of cell preparation events normalized by the number of UEs per minute  for three different LLM cases. As shown in the figure, the number of cell preparation events is reduced with LLM-F by $5\%$ percent compared to LLM without filtering.  Fig.~\ref{fig:resourcereservation_ds_llm} shows in bar the resource reservation percentage of LLM when filtering and dynamic switching is enabled. It is observed that applying L2 filtering at the network increases the network resource reservation by $23\%$ compared to LLM. This is because the delay caused by L2 filtering leads the network to reserve the resources prepared for the UE longer compared to LLM which results in higher resource reservation percentage. Moreover, the cumulative distribution function (CDF) of the resource reservation duration in a cell is shown in Fig.~\ref{fig:resourcereservation_ds_cdf}  for different LLM procedures. According to the figure, the resource reservation duration in a prepared target cell is slightly higher at $50\text{th}$ and $90\text{th}$ percentiles for LLM-F compared to LLM.

As for the impact of dynamic switching on LLM,  the reliability metric of LLM-F-DS is similar to that of LLM-F as shown in Fig.~\ref{fig:kpi_comparison}. Fig.~\ref{fig:sentcellprep} shows that the number of cell preparation events is reduced by $7\%$ in LLM-F-DS compared to LLM-F. This gain is achieved due to the retention of the prepared target configurations in dynamic switching. However, as the prepared cells are not released after each handover, the network resource reservation is increased by around $200\%$ against LLM-F as shown in Fig.~\ref{fig:resourcereservation_ds_llm}. This means that in $43\%$ of the simulation duration a resource is reserved for a UE by the network. The CDF of the resource reservation duration in Fig.~\ref{fig:resourcereservation_ds_cdf} shows that at the $95\text{th}$ percentile, the resource reservation of LLM-F-DS  is twice as higher than that of LLM-F. The difference gets even higher in higher percentiles of the CDF.

\section{Conclusion}
In this paper, we have evaluated the mobility performance of both lower and higher layer mobility procedures that are currently being discussed in 3GPP. According to our analysis, Lower Layer Mobility outperforms Higher Layer Mobility with respect to radio link reliability. The number of ping-pongs in LLM is much higher than in higher layer mobility, however as the handover interruption time is small, i.e., $1/80$ of higher layer mobility, the impact of ping-pongs on mobility performance is not critical. Nevertheless, ping-pongs increase the network and radio interface signalling overhead. Moreover, it has been shown that the introduction of L2 filtering at the network decreases the number of ping-pongs in LLM by $28\%$ without degrading reliability.


In addition, 
dynamic switching reduces the cell preparation events in the network for LLM and thus decreases the signaling overhead in network and radio interfaces. However, as the network continues to reserve resources for the UEs after each handover in dynamic switching, the network resource reservation overhead is significantly increased.

The performance of LLM can be further improved by optimizing the parameters controlling the handover preparation and execution. This can be achieved by using advanced AI/ML techniques which leverage the past experiences of the UEs in the network.


%



\bibliographystyle{IEEEtran}
\bibliography{main.bib}

\begin{thebibliography}{10}
\providecommand{\url}[1]{#1}
\csname url@samestyle\endcsname
\providecommand{\newblock}{\relax}
\providecommand{\bibinfo}[2]{#2}
\providecommand{\BIBentrySTDinterwordspacing}{\spaceskip=0pt\relax}
\providecommand{\BIBentryALTinterwordstretchfactor}{4}
\providecommand{\BIBentryALTinterwordspacing}{\spaceskip=\fontdimen2\font plus
\BIBentryALTinterwordstretchfactor\fontdimen3\font minus
  \fontdimen4\font\relax}
\providecommand{\BIBforeignlanguage}[2]{{%
\expandafter\ifx\csname l@#1\endcsname\relax
\typeout{** WARNING: IEEEtran.bst: No hyphenation pattern has been}%
\typeout{** loaded for the language `#1'. Using the pattern for}%
\typeout{** the default language instead.}%
\else
\language=\csname l@#1\endcsname
\fi
#2}}
\providecommand{\BIBdecl}{\relax}
\BIBdecl

\bibitem{3gpp38913}
3GPP, ``{Study on Scenarios and Requirements for Next Generation Access
  Technologies},'' {3rd Generation Partnership Project}, TR {38.913}, Mar.
  2022.

\bibitem{umurphd}
U.~Karabulut, ``Mobility management in 5g beamformed systems,'' Ph.D.
  dissertation, Technische Universität Dresden, 2021.

\bibitem{R21706489}
A.-L. S.~B. Nokia, ``{Conditional handover basic aspects and feasibility in
  Rel-15, TSG-RAN WG2 NR Adhoc Meeting \#2},'' {3rd Generation Partnership
  Project}, Tech. Rep. {R2-1706489}, Jun. 2017.

\bibitem{9824571}
J.~Stanczak, U.~Karabulut, and A.~Awada, ``Conditional handover in 5g -
  principles, future use cases and fr2 performance,'' in \emph{2022
  International Wireless Communications and Mobile Computing (IWCMC)}, 2022,
  pp. 660--665.

\bibitem{3gpp38133}
3GPP, ``{NR; Requirements for support of radio resource management },'' {3rd
  Generation Partnership Project}, TS {38.133}, Jun. 2022.

\bibitem{RP193133}
------, ``{Further enhancements on MIMO for NR},'' {3rd Generation Partnership
  Project}, 3GPP™ Work Item Description {RP-193133}, Dec. 2019.

\bibitem{RP212535}
------, ``{Further enhancements on MIMO for NR},'' {3rd Generation Partnership
  Project}, 3GPP Work Item Description {RP-212535}, Sep. 2021.

\bibitem{3gppevm}
------, ``{Moderator summary for multi-beam enhancement: EVM, 3GPP TSG RAN WG1
  \#102-e e-Meeting},'' {3rd Generation Partnership Project}, Decision
  {R1-2007151}, Aug. 2020.

\bibitem{rp221799}
M.~Inc., ``{Revised WID on Further NR mobility enhancements, TSG-RAN Meeting
  \#96},'' {3rd Generation Partnership Project}, Tech. Rep. {RP-221799}, Jun.
  2022.

\bibitem{nokiawp}
\BIBentryALTinterwordspacing
Nokia, ``{White paper: Rock solid mobility innovations from 5G to
  5G-Advanced},'' {Nokia}, White paper, Jun. 2022. [Online]. Available:
  \url{https://onestore.nokia.com/asset/212564}
\BIBentrySTDinterwordspacing

\bibitem{3gpp38331}
3GPP, ``{NR; Radio Resource Control (RRC); Protocol specification },'' {3rd
  Generation Partnership Project}, TS {38.331}, Jul. 2020.

\bibitem{3gpp38901}
------, ``{5G; Study on channel model for frequencies from 0.5 to 100 GHz},''
  {3rd Generation Partnership Project}, TR {38.901}, Dec. 2020.

\end{thebibliography}

\end{document}